\documentclass[]{spie}  

\usepackage{amsmath,amsfonts,amssymb}
\usepackage{graphicx}
\usepackage[colorlinks=true, allcolors=blue]{hyperref}
\title{ Generation and Measurement of Ultrashort Free Electron Laser Pulses in Ultraviolet and Soft-X-ray Spectral Regions }
\author[a]{N. S. Mirian}

\affil[a]{Deutsches Elektronen-Synchrotron DESY, Notkestraße 85, 22607 Hamburg, Germany}
\affil[a]{formerly at Elettra-Sincrotrone Trieste S.C.p.A., 34149 Trieste, Italy}

\authorinfo{Najmeh Mirian: najmeh.mirian@desy.de}
\begin{document} 
\maketitle
\begin{abstract}
Over the last few years, tremendous progress has been gained in the generation and application of ultrashort radiation pulses. Recently, free-electron lasers generating ultrashort pulses with high peak power from the
extreme ultraviolet (EUV) to the soft-X-ray region are opening a wide
range of new scientific opportunities. Taking advantage of this short timescale permits probing ultrafast, out-of-equilibrium dynamics and the high intensities are key for nonlinear optics. 
The core structure of the extremely important light elements carbon, nitrogen, and oxygen can be accessed by soft-X-ray wavelengths by providing chemical sensitivity.
Externally seeded free-electron lasers generate coherent pulses with the ability to be synchronized with femtosecond accuracy. In this contribution, we present new achievements in the generation of coherent ultrashort pulses in the range of EUV to the soft X-ray in externally seeded FELs. In particular, we present the recently successful robust experiment at FERMI in Trieste, where few-femtosecond extreme-ultraviolet pulses were generated and characterized in terms of energy, and duration via autocorrelation. 
\end{abstract}
\keywords{Attosecond pulse, femtosecond pule, free electron laser, ultrashort
pulse, seeded FEL}

\section{INTRODUCTION}

The study of the temporal evolution of matter down to the sub-femtosecond time scale and atomic space scale is essential for chemical, physical and biological processes. 
In this respect, free-electron lasers (FELs) open a new door to a wide range of new scientific opportunities by providing unprecedented pulse properties in the soft and hard-X-ray spectral ranges, in terms of brilliance, tuneability, intensity, coherence, and last but not least, pulse duration.  
Intense ultrashort FEL pulses have paved the way for time resolution and for the observation of nonlinear and ultrafast processes in the study of the interaction
of radiation with a matter involving shallow and deep-core electron levels. 

In general, FEL sources can be operated in several different ways. To date, most
of the existing short-wavelength FELs such as FLASH \cite{rossbach201910}, LCLS \cite{lcls}, and EuXFEL \cite{decking2020mhz} are operated in the self-amplification of spontaneous emission (SASE) \cite{PhysRevLett.57.1871}. Though SASE FELs are able to generate extremely high brilliance, the temporal structure of the pulses includes a series of micro-pulses that individually have random phases and highly fluctuating peak intensity and time duration. Synchronization of SASE devices to external sources is normally limited by the temporal jitter of the electron accelerator. 
Ultrashort pulses in FELs can be generated in SASE mode. In SASE FELs the most straightforward method consists in shortening the pulse region that has the right properties to lase \cite{Emma2004}. Several other schemes were proposed to reduce the pulse duration at the femtosecond level or shorter.~\cite{Marinelli2016,Behrens2014,Ding2015,Zholents2005,Ding2016,Huang2017,Marinelli2017} 

An alternative operation mode for an FEL is based upon \textquoteleft seeding\textquoteright{} techniques. When the FEL operation wavelength is in the VUV to soft X-ray spectral range,  the FEL amplification can be initiated by a coherent radiation pulse generated by an external laser, the seed. Seeded FELs produce output pulses with a well-defined temporal shape and intensity stability. The availability of a seed of sufficient intensity becomes increasingly critical as the wavelength decreases. In the hard X-rays, the only seed available with sufficient intensity consists of the radiation emitted by the beam itself, in a self-seeded scheme \cite{self-seeding}
In externally seeded FELs, before emitting coherent radiation, the electrons interact with an external coherent source (the seed, generally at UV wavelengths) in a modulator, and then under given conditions, one of the higher order harmonics of the seed is selected, and amplified in a long amplifier, named radiator. 

\begin{figure}
\begin{center}
\includegraphics[width=9cm]{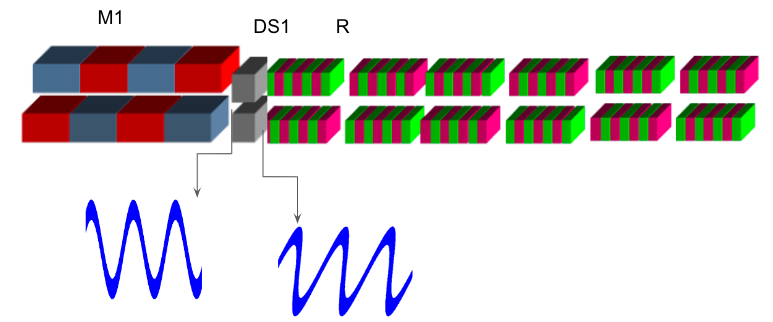}\caption{HGHG scheme \label{fig:HGHG-scheme}}
\end{center}
\end{figure}

 Two well-known seeded FEL schemes are the high-gain harmonic generation scheme (HGHG) \cite{PhysRevA.44.5178,Allaria_2010} and the echo enable harmonic generation (EEHG).\cite{primoz2019,Xiang2009}
Fig.~\ref{fig:HGHG-scheme} shows  the HGHG scheme. The electron beam after being accelerated in a linear accelerator interacts with an external coherent laser in UV wavelengths at a modulator (M1). The laser
produces a sinusoidal energy modulation in the electron beam. The beam then goes through the longitudinally dispersive section (DS1). The dispersive section creates high harmonic bunching at short wavelengths in the current distribution. The beam then enters the radiator (R) and emits coherent light and exponentially amplifies it. HGHG scheme is able to produce bunching in the electron beam down to the XUV region. 

\begin{figure}
\begin{center}
\includegraphics[width=10cm]{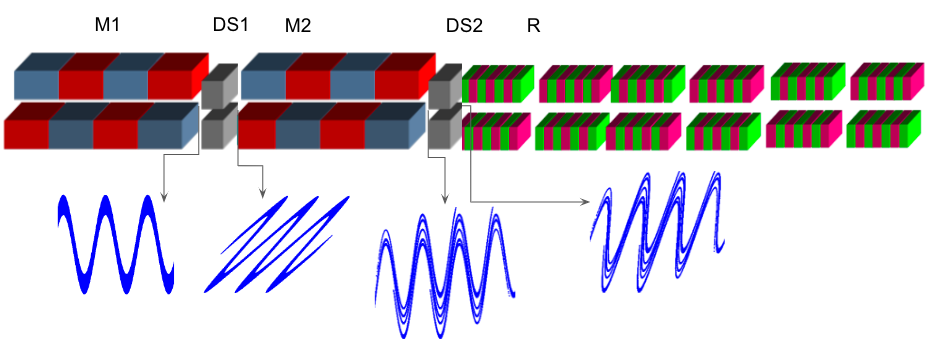}\caption{EEHG scheme \label{fig:EEHG-scheme}}
\end{center}
\end{figure}

Fig.~\ref{fig:EEHG-scheme} presents the EEHG scheme. EEHG employs two setups of external laser and dispersive section to efficiently generate bunching in the electron beam down to soft X-rays. The first laser produces a sinusoidal energy modulation in the electron beam at the first modulator (M1). Then the beam travels through a strong longitudinally
dispersive section (DS1) that folds over the sinusoids and produces
 filaments  in the beam phase space distribution.
The second laser then modulates the filaments' distribution
at the second modulator (M2). Finally the second dispersive section (DS2)
generates high harmonic bunching at short wavelengths in the current
distribution \cite{Stupakov2009,Xiang2009,primoz2019}.

FERMI \cite{Allaria_2010} is an externally seeded FEL user
facility, based on the high-gain harmonic generation scheme (HGHG) \cite{PhysRevA.44.5178}. FERMI has two FEL beamlines, FEL-1 is single-stage HGHG \cite{Allaria2012}and FEL-2 is based on two-stage HGHG, separated by magnetic delay line to facilitate the fresh bunch technique \cite{Allaria2013}. 
Regarding the generation and measurement of an ultrashort pulse in a short-wavelength range at seeded FEL, several
methods based on manipulation of the electron beam, altering seed laser parameters, using the FEL amplification concept, etc, have been proposed and demonstrated by the FERMI team in recent years.
In this contribution, we summarize the recent FERMI team activities with respect to the first extended study of ultrashort seeded FEL pulse generation and of the seeded FEL pulse temporal properties characterization.
\begin{figure}
\begin{center}
\includegraphics[width=10cm]{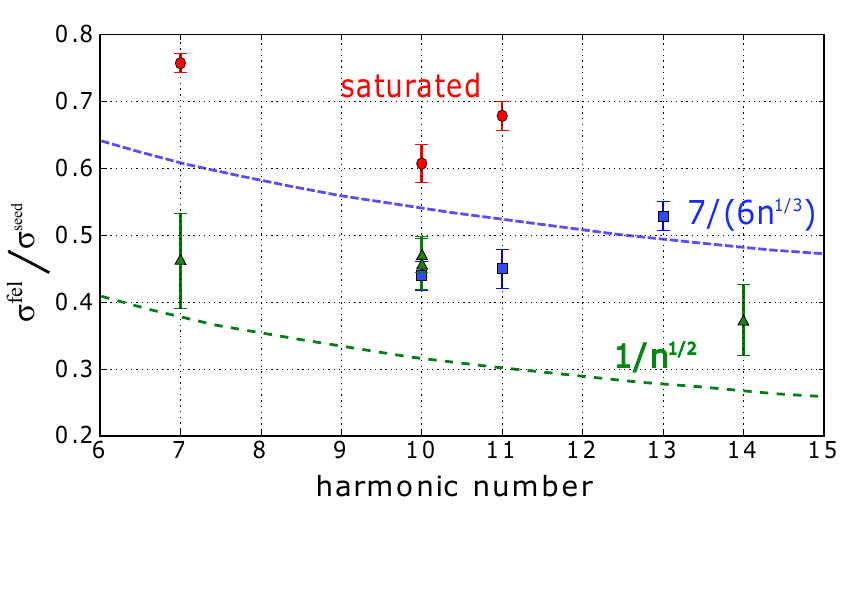}\caption{The FEL rms pulse duration versus harmonic number. Blue squares represent
two-color gas photo-emission cross-correlation method results, green triangles
show high-resolution single-shot solid-state cross-correlation method
results. The Green dashed line shows the function $1/\sqrt{n}$, where n
is the harmonic number. The blue dotted line shows the optimized peak bunching
factor estimated by theory, scaling as about $7/(6n^{1/3})$. The
red dots are larger duration pulses obtained with an increased
seed power or dispersion. (see Fig.11 in Ref.~\citenum{PhysRevX.7.021043})
\label{fig:PRXfig}}
\end{center}
\end{figure}

\section{Measurement and estimation of seeded FEL pulse}

Characterization of the temporal intensity profile of free-electron laser (FEL) pulses and determination of the pulse duration are extremely
crucial for exploring the new perspectives suggested by FEL sources. The analysis of experimental data, particularly
for nonlinear interactions requires accurate knowledge of pulse peak
power in the time domain. In the last decade, several direct and indirect
methods have been developed to illustrate reliable temporal profiles,
both on average and on a single-shot basis. S. D\"usterer
and his colleagues in DESY utilized and compared nine different pulse
duration measurement methods to characterize the photon pulse duration,
generated in the wavelength range of 13.5 to 27 nm at FLASH, the self-amplified
spontaneous emission FEL at DESY in Hamburg \cite{PhysRevSTAB.17.120702,Günther2011}.
They represented autocorrelation with gas phase and semiconductors
detectors, and cross-correlation with terahertz laser (for comparison
see Fig.12 in Ref.~\citenum{PhysRevSTAB.17.120702}). 

FERMI team has measured the pulse shape of an extreme ultraviolet
and soft X-ray externally seeded FEL operating in high-gain harmonic
generation mode by different methods \cite{PhysRevX.7.021043, DeNinno2015, Mahieu}. 
They characterized a single-shot spectrotemporal of the seeded FEL
pulses in a specific double-pulse configuration by means of the spectral phase interferometry for direct electric field reconstruction (SPIDER)
technique\cite{Mahieu}. This method is able to show the
reconstruction of the amplitude and phase of the pulses. However,
SPIDER method is applicable to a specific machine setup since it imposes
several constraints on the FEL operation. 

In 2017, the FERMI team presented two
different cross-correlation methods: two-color photo-emission cross-correlation
(method A, Sec. III A in Ref.~\citenum{PhysRevX.7.021043}) and high-resolution single-shot solid-state cross-correlation (method B, Sec. III B in Ref.~\citenum{PhysRevX.7.021043}). Basically in both cases, the FERMI team measured
the similarity of the unknown FEL pulse with an external optical laser.
In that work, the FEL temporal pulse measurement was carried out in a range of FEL wavelengths, as different harmonics of the
seed laser and machine settings, and compared to the predictions of
a theoretical model. Blue squares in Fig.~\ref{fig:PRXfig}) refer to measured rms pulse duration of different harmonic FEL amplification by the two-color gas photo-emission cross-correlation method and green triangles
indicate the measurement results by high-resolution single-shot solid-state cross-correlation method.
The results obtained with the two methods were in agreement with the proposed
theory. In the theory, presented in that paper, the pulse duration
in seeded FEL performance is estimated by the bunching factor parameter, the Fourier coefficient of the electron density modulation at the resonant frequency of amplification. The amplitude of the bunching factor
in the time domain estimates the rms FEL pulse duration under the conditions
of low seed power and low dispersion, which should be around $\sigma_{s}/\sqrt{n},$
where n is the number of seed harmonics and $\sigma_{s}$ is the rms
value of the seed laser. This function is shown by the Green dashed line in Fig.~\ref{fig:PRXfig}.  As well, it was proved that in the case of optimal seed power and dispersion, the rms value of the FEL pulse duration is approximated
by the function $7\sigma_{s}/(6n^{1/3})$, presented by the dashed blue line in Fig.~\ref{fig:PRXfig}.

\section{femtosecond shaping of seeded FEL pulse by chirped seed}

Though ultrafast pulse shaping techniques are well demonstrated in
the infrared and visible, direct application of such methods to an
X-ray FEL is not straightforward. In 2016, K. C. Prince et al.
showed the necessity of controlling and shaping the FEL spectral and temporal
in Ref.~\citenum{Prince}. The ability to control and shape the spectral
and temporal of extreme-ultraviolet and soft X-ray pulses produced
by seeded FEL at the FERMI facility, has been demonstrated by employing
the precise manipulation of the linear frequency chirp and intensity of the seed
laser pulse as well the variation of longitudinal dispersion.  \cite{PhysRevLett.115.114801,PhysRevAccelBeams.23.060701}
In this section, we discuss the experiments in HGHG and EEHG performances,
showing femtosecond shaping of seeded FEL pulse. 
\subsection{HGHG}
In general, the length of seeded FEL pulses is determined by the length of the seed laser and is typically in the range between a few tens of fs and 100 fs. Due to several mechanisms in the seeding process, shorter pulses (in the range of a few fs) are not directly accessible with HGHG. However, fs time resolution experiments can still be done taking advantage of the high degree of longitudinal coherence of the produced pulses. Indeed, if the length of the pulse is limited to tens of fs the phase along the pulse can be controlled with much higher accuracy.

Exploiting the perfect phase locking between two harmonics of the same seed laser makes it possible to perform coherent control experiments with attosecond accuracy \cite{Prince}. A proper combination of 3 and 4 harmonics of the same seed was used for the generation of an attosecond pulse train in the UV, presented by P.K. Maroju and the FERMI team in Ref.~\citenum{Maroju2020}. All these methods for accessing the attosecond time scale with tens of fs pulses rely on the accurate control of the phase which is possible thanks to the presence of an optical laser as the seed  starting the emission process.
\begin{figure}
\begin{center}
\includegraphics[width=12cm, height=7cm]{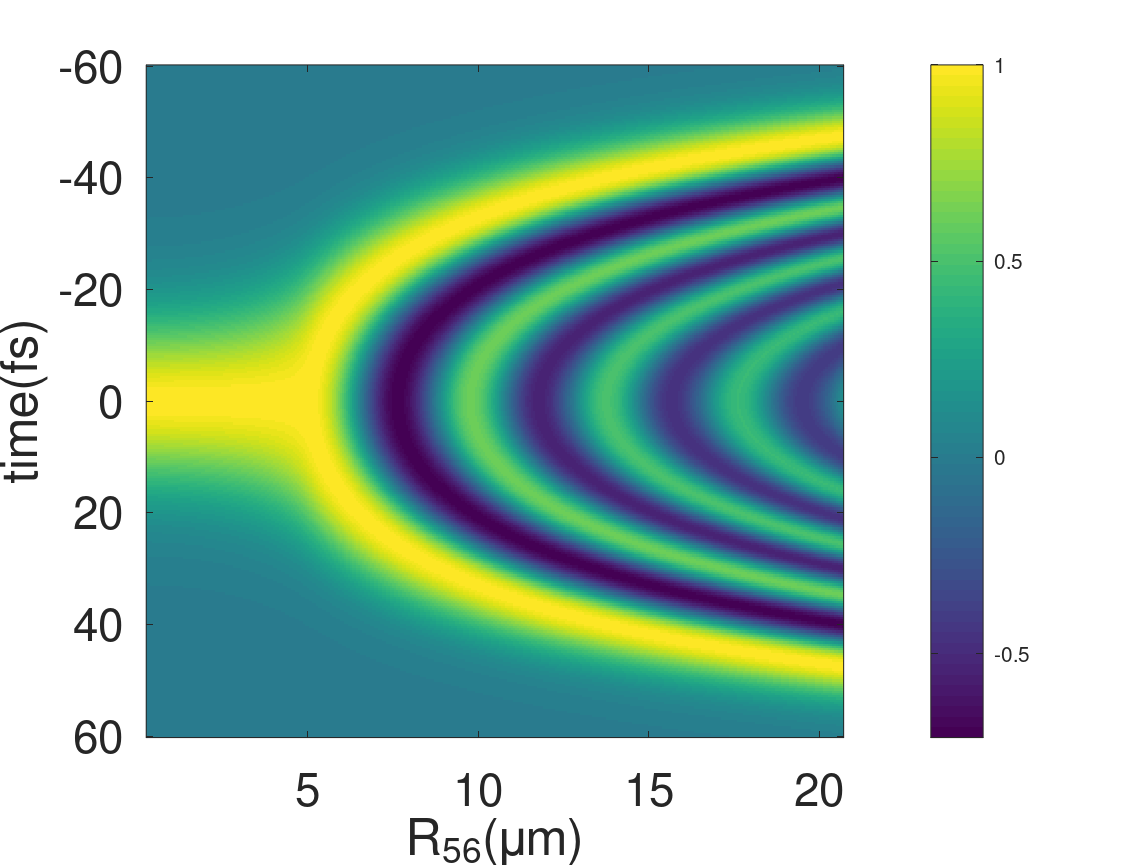}\caption{Normalized FEL pulse profile, simulated from bunching factor function versus longitudinal dispersion $B=R_{56}$. The FWHM seed pulse duration was assumed 50 fs and the FEL wavelength is the 7th harmonic of the seed laser (seed wavelength is 260 nm ).   \label{fig:fs_shaping_HGHG}}
\end{center}
\end{figure}

Dedicated experiments at FERMI showed in Ref. \citenum{PhysRevLett.115.114801} that proper control of the seed laser and electron beam parameters allows controlling the phase profile in the FEL pulse.
Thanks to the overbunching that can be created with a proper adjustment of the seed laser intensity and longitudinal dispersion ($R_{56}$), it is possible to induce  a splitting of the FEL pulse. This allows generating of two (or eventually more pulses) in the time domain. Suche a structure is also visible in the spectral domain. This fact is presented in Fig.~\ref{fig:fs_shaping_HGHG}. 
We estimated the shape of the FEL pulse bunching factor profile in the time domain. Fig.~\ref{fig:fs_shaping_HGHG} shows the map of the FEL pulse profile as a function of longitudinal dispersion of the first chicane in the HGHG scheme (Ds1 in Fig.~\ref{fig:HGHG-scheme}). 
At a certain point of dispersion, the FEL pulse splits into two or more femtosecond pulses. The time separation of the pulses can be controlled by the chirp and intensity parameters of the seed laser pulse. More information about this technique is reported in Refs.~\citenum{PhysRevLett.115.114801, PhysRevLett.110.064801}.
In addition, if two phase-locked pulses are used for seeding instead, it has been shown that a phase-locked FEL pulse can be generated \cite{PhysRevLett.116.024801}. This again allows accessing the sub fs time resolutions simply controlling the phase between the two seed laser pulses.
\subsection{EEHG}

\begin{figure}
\begin{center}
\includegraphics[width=11cm]{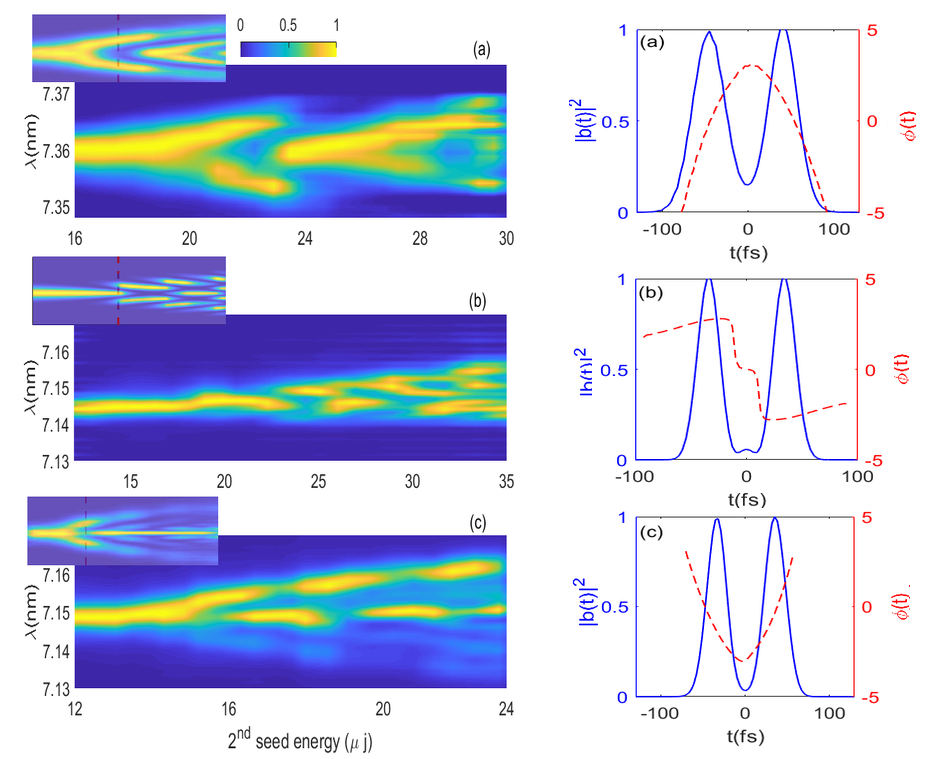}

\caption{Left: Evolution of the spectral shape as a function of second seed laser
energy with three different circumstances of the linear frequency
chirp on it. (a): $(GDD=-3500\,fs^{2})$ (b): $(GDD=0\,fs^{2}),$
(c):$(GDD=+2300\,fs^{2})$ . Insets are theoretical and simulated
spectral maps reproduced in the spectral domain.
In all cases, the e-beam energy, energy spread, current, and emittance
were \ensuremath{\simeq}1.3 GeV, \ensuremath{\simeq}270 keV, 700
A, and \ensuremath{\simeq}1 mm mrad, respectively. For (a) and (b),
the e-beam (positive) quadratic energy curvature was \ensuremath{\simeq}8
MeV/ps2. The maps in (a) and (b) consist of 20 accumulated single-shot
spectra at harmonic 37 for each value of second seed energy, and data
in (c) were taken 50 shots pf spectra at harmonic 36. First seed energy
is $52\,\mu j$ ($A_{1}=4$ in simulation), $B_{1}=11.4$ and $B_{2}$
was in optimal value.
Right: Time-domain normalized pulse
shape (blue solid line) and phase (dashed red line) of the
 time profile of the bunching factor, obtained from the Fourier transform of the
(calculated) bunching spectra indicated along the dashed lines in the insets of
right figures for (a) $GDD=-3500\:fs^{2}$, 
(b) $GDD=0\,fs^{2}$, (c) $GDD=2300\:fs^{2}$.\label{fig:fs_shaping_EEHG}}
\end{center}
\end{figure}

Echo enable harmonic generation FEL performance was executed at FERMI
in 2018 for half a year. FERMI demonstrated the first high-gain lasing of
an EEHG FEL in the soft X-ray region at 7.3 nm 
and 5.9  nm, which
was the thirty-sixth and forty-fifth harmonics of the seed wavelength
\cite{primoz2019}. The FERMI team prepared the EEHG setup by altering the
FEL-2 line which applies the double-stage HGHG approach to extend the
harmonic range up to 60 -70 \cite{Allaria2013}. In following the
EEHG demonstration, N. S. Mirian and FERMI team in Ref. ~\citenum{PhysRevAccelBeams.23.060701}
illustrated that spectral and temporal control over the FEL pulse
can be achieved by tuning the second seed laser parameters. In the EEHG
setup, the second seed laser directly corresponds to the intensity modulations
in the temporal domain. Increasing second seed power develops the FEL spectrum
intensity modulations. Fig.~\ref{fig:fs_shaping_EEHG}-left shows the
spectral maps (spectrum vs second seed laser energy) for different
values of the group delay dispersion (GDD) of the second seed laser; the insets are theoretical spectral maps obtained from EEHG equations
(see Eq. (2) in Ref.~\citenum{PhysRevAccelBeams.23.060701} ) and
agree relatively well with the experiment. Fig.~\ref{fig:fs_shaping_EEHG}-right represents the calculated temporal domain reconstruction and phase of the FEL pulses for the case of the spectrum indicated by the vertical dashed line in the inset figures. 
In Fig.~\ref{fig:fs_shaping_EEHG}(a),
we plot the results for a relatively high (absolute) value of the
GDD $\sim$ \textminus 4100 $fs^{2}$. Starting from the optimum energy modulation that maximizes bunching, increasing the seed intensity, the spectrum initially characterized by a single peak broadens until it develops intensity modulations as a function of the wavelength due to the process of electron overbunching and rebunching \cite{PhysRevLett.115.114801}. For such high GDD, this intensity modulation directly corresponds to the intensity modulation in the temporal domain due to the spectrotemporal
equivalence. A split spectrum in the wavelength domain indicated by the vertical dashed line in the inset of Fig.~\ref{fig:fs_shaping_EEHG}(a)-left, thus corresponds to few-femtosecond double-pulse structure with an overall negative quadratic chirp in the temporal domain, as shown in Fig.~\ref{fig:fs_shaping_EEHG}(a)-right. The temporal separation between the pulses is controlled by the GDD value. The spectral map for zero GDD, Fig.~\ref{fig:fs_shaping_EEHG}(b)-left, corresponds to a  Fourier transform limited FEL pulse
structure with a flat phase within the individual spikes, as demonstrated in Fig.~\ref{fig:fs_shaping_EEHG}(b)-right for the case of the spectrum indicated
by the vertical dashed line in the inset of Fig.~\ref{fig:fs_shaping_EEHG}(b)-left.
The spectral map for the case of a positive GDD of 2000 $fs^2$, is represented in Fig.~\ref{fig:fs_shaping_EEHG}(c)-left and reconstruction of the phase and temporal profile of the FEL pulse for the case of the spectrum indicated by the vertical dashed line in the inset, are shown in Fig.~\ref{fig:fs_shaping_EEHG}(c)-right. 
This result demonstrates deterministic control of the FEL
phase in time and in the spectrum domain by changing the parameters of
second seed laser in EEHG setup. 

\section{Chirped pulse amplification in free-electron laser}
Chirped pulse amplification (CPA) in optical lasers is known as a revolutionary
technique. CPA approach allows the amplification of the ultrashort laser
pulse up to the petawatt level, by stretching out the laser pulse
temporally and spectrally first, then amplifying, and compressing the pulse at
the end. The stretching and compression applied by the devices
with the concept that the different color components of the pulse
travel different distances. The first femtosecond FEL by CPA was proposed
by L.H. Yu et al. in 1994 \cite{PhysRevE.49.4480}. Like
the classical lasers, the CPA technique at seeded FEL relies on stretching
the seed pulse in time by means of a linear frequency chirp before
amplification, that enables us to extract energy from the whole electron
bunch and enhances the FEL pulse energy at saturation. To sustain
the FEL resonant condition, the large bandwidth of the seed laser is
matched by proper chirp in the energy of the electron beam. 

D. Gauthier with the FERMI team in Ref.~\citenum{CPA-FEL} presented the
experimental implementation of chirped pulse amplification on HGHG
performance in the extreme ultraviolet wavelength range.
They imposed a positive linear frequency chirp on the seed pulse and
stretched it up to 290 fs. The chirped seed manipulated the chirped electron beam in the modulator (M1 in Fig.~\ref{fig:HGHG-scheme}). Then, harmonic 7 of the seed
was chosen for amplification in the radiator (R in Fig.~\ref{fig:HGHG-scheme}). The final FEL pulses before and after
compression were characterized during the experiment. The temporal
characterization was done by employing the cross-correlation scheme,
based on the ionization of a Helium gas sample by the FEL pulse in the
field of an intense infrared laser pulse with 90 fs FWHM
duration. The technical detail of measurement and compression of the
FEL pulse was reported in Ref.~\citenum{CPA-FEL}. Fig.3 of 
that reference summarized the results of the experiment. The average FWHM spectral width was $4.46\times10^{-2}$ nm,
while fluctuations between consecutive measurements were in order of
few percent. Three independently measured cross-correlation curves,
associated with the second sideband of the photo-electron spectrum,
were presented for the non-compressed case
and for the maximum compression cases. 
After reconstruction of the FEL pulses from cross-correlation measurements,  the obtained FWHM pulse duration for the case of no-compression
was \ensuremath{\sim}143 fs (with fluctuations of few
percent). And after compression, the FWHM pulse duration
was 40 fs which was a significant shortening of the FEL pulse compared with the no-CPA case. 
\\
It is theoretically possible to implement this technique into the soft x-ray spectral window with a pulse duration much below femtosecond, however, the technique suffers from the limited transmission of the grating system required to compress the pulses that at short wavelength can be lower than 1 percent.

The CPA performance result showed that CPA cannot only compensate for the linear frequency chirp induced by seed control in the FEL pulse, but also unwanted residuals generated by other sources, such as the quadratic curvature of the electron-beam energy profile, seed transport, and so on. 
\begin{figure}[tbh]
\includegraphics[width=0.98\textwidth]{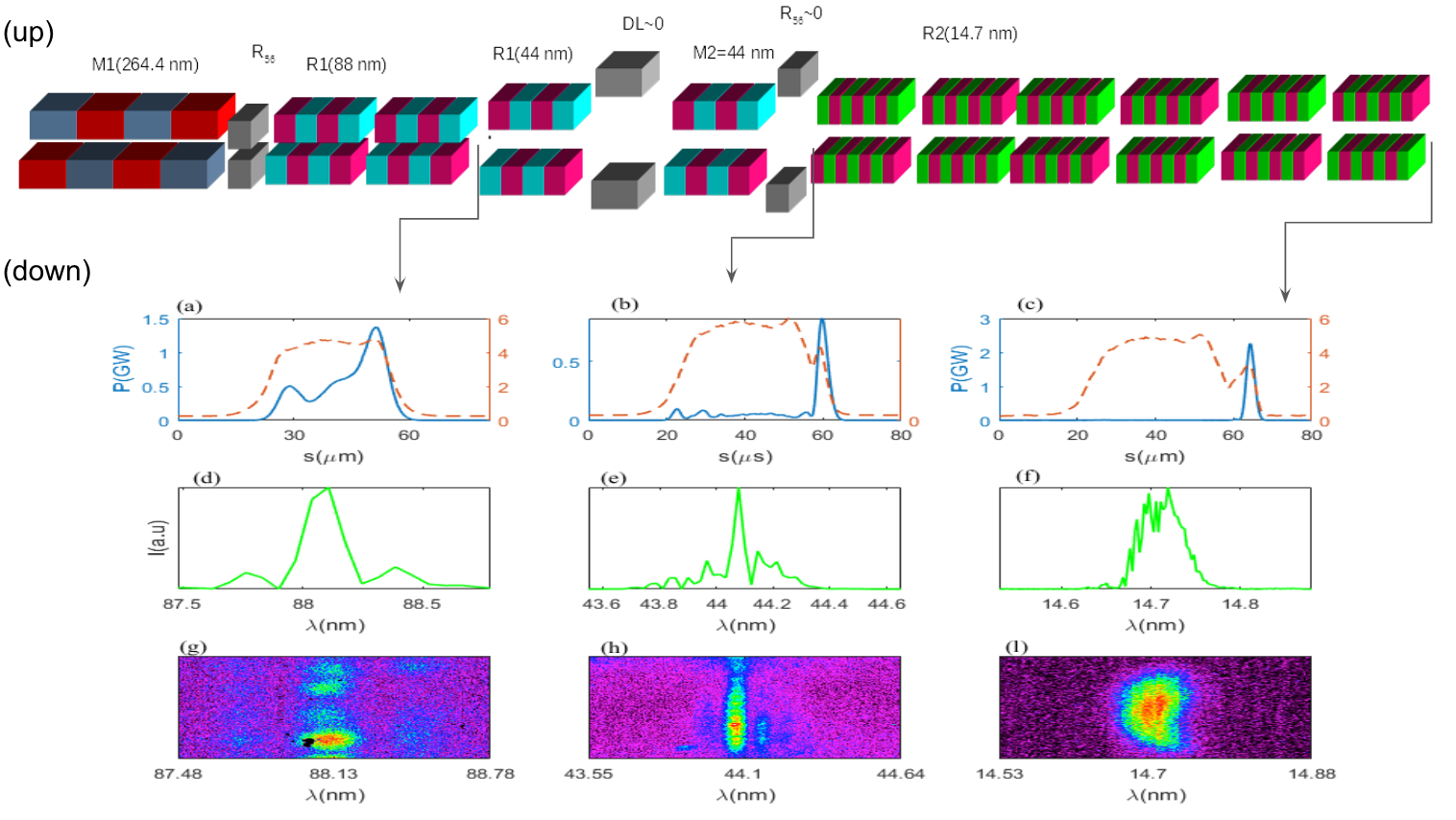}
\caption{\label{fig:SR} Up: the cascade superradiant scheme. The FEL-2 line was divided into three sections. The electron beam was modulated with the coherent seed at 264.4 nm wavelength in modulator, M1.  With the help of the longitudinal dispersion of the first chicane, we selected the third harmonic and amplified it in two undulator sections. In the next two undulator sections, we tuned the resonance on the 6th harmonics of the seed laser (44 nm), and then in the last radiator (R2) included six undulator sections, the harmonic 18th was amplified. The longitudinal dispersion of the Delay line (DL)  and second dispersion section was set to zero. 
Down: Simulated radiation pulse (blue solid line) and e-beam energy spread profile (red dashed line) after (a)
first, (b) second, and (c) third stage. Simulated spectrum profile (green solid line) at (d)
first, (e) second, and (f) third stage. Measured spectrum after (g) first, (h) second, and (l) third stage.}

\end{figure}
\section{ superradiant spike in free-electron laser amplifier}
Nonlinear FEL evolution and pulse propagation effects are important and interesting methods to shorten the FEL pulse. Bonifacio and his collaborators \cite{Bonifacio1985, Bonifacio1990b} elaborated a regime where FEL evolution is dominated by the nonlinear effects  associated  with  saturation  and  by  pulse propagation effects associated with  slippage. This regime is characterized by the propagation of a solitary wave-like self-similar pulse  with peak power scaling as $N^2$, where N is the number of electrons\cite{DiCke}.
Amplification of a soliton, superradiant is known as a popular and robust method to generate ultrashort intense FEL pulse. The interesting advantage of the superradiant approach with respect to the other methods is that it can exploit the FEL dynamic process to beat the gain bandwidth limit. FEL enters this regime when a short pulse grows up to saturation while it is advancing on the electron that is delayed by the wiggling motion in the undulator. The pulse peak power continues to grow proportionally to the square of the distance covered along the undulator and the pulse duration is proportional to the inverse of the square root of the distance. \\
The propagation of the superradiant pulse is performed by the amplification of harmonic emission and enhancement of the harmonic bunching on the head of the pulse. This harmonic bunching can be used in a second radiator section, tuned at one harmonic of the seed. This approach enables us to produce a superradiant pulse at shorter wavelengths \cite{Giannessi2005}. The pulse shortening via superradiance in a single stage was demonstrated in the infrared (800 nm) \cite{Watanabe2007}. The possibility of having superradiant propagation in a two-stage cascade was verified at SPARC \cite{Giannessi2012}. In that case, the two stages were tuned at 400 nm at 200 nm respectively.\\
In 2020, N. S. Mirian and FERMI team demonstrated the generation of few-femtosecond extreme-ultraviolet pulses via a multi-stage superradiant approach and measured the pulse by autocorrelation method assisted by photo-emission of Argon gas \cite{Mirian2021}. The FERMI team achieved pulses four times shorter, with higher peak power, than in the standard high-gain harmonic generation mode and proved that the pulse duration matched the Fourier transform limit of the spectral intensity distribution. \\
The experiment was accomplished at 
the FERMI FEL-2 line. The FERMI FEL-2 beamline consists of 13 undulators of 
three different types. FEL-2 beamline consists of a two-stage HGHG scheme, separated by a magnetic delay line. The magnetic delay line is considered to make the fresh technique possible such that by delay it can change the lasing window from the tail of the electron beam to the head. 

Fig.~\ref{fig:SR} summarizes the method and results of the superradiant experiment at FERMI. Fig.~\ref{fig:SR}(up) represents the three-stage layout of the experiment. In the first modulator (M1) the seed interacted with the electron beam and induced an energy modulation on it in the range of seed wavelength (264.4 nm). This energy modulation was converted into the density modulation with components at all the harmonics of the seed in the first dispersive section (DC1), and the first two undulator 
sections (R1) were tuned into an additional frequency-conversion stage
at 88 nm (third harmonic of seed laser), to initiate the cascade with a shorter gain length and longer slippage; then the next two undulator sections (R2) were tuned to 44 nm radiation wavelength. The delay line (DL) was used as a phase shifter to match the phase between the two undulator sections. The dispersive section (DS2) was set to zero. The second-stage radiator containing a chain of six undulators produced the FEL radiation short pulse in 14 nm wavelength. 
We showed the evolution of pulse in the three stages of the cascade both in Genesis \cite{genesis} simulation and from the spectral measurements done during the experiment with the spectrometer in PADRES (Photon Analysis, Delivery, and Reduction System) \cite{RSI_PADReS, JSR_PADReS}.
The first stage at 88 nm was added to start the cascade with higher gain (shorter gain length) and longer slippage in the first sections. In this case, it was possible to choose the right value of the seed peak power and of the first dispersive section, to  reach the saturation in the second undulator. Fig.~\ref{fig:SR}(a) shows the power profile (blue line) of the pulse and the slice energy spread (red line) along the electron beam at the end of the second undulator as obtained in a genesis simulation. We see that the pulse started to evolve in the superradiant regime with a spike in the head growing above saturation. The spectrum at the end of the second undulator in that stage consisted of two symmetric sidebands that showed the distortion of the longitudinal profile. We see a good agreement between simulation (Fig.~\ref{fig:SR}(d)) and measurement (Fig.~\ref{fig:SR}(g)) in regard to the shape of the spectrum. The saturation was associated with an increase in the energy spread on the region of the beam, interacting with the radiation.  In the second stage, the bunching at the second harmonic started to radiate at 44 nm. This emission was stronger on the head where the bunching was higher and the energy spread was lower. On the contrary, the radiation was strongly reduced in the tail by the energy spread accumulated in the previous stage. This is shown in Fig.~\ref{fig:SR}(b).  Even at this stage, there was a good agreement between the simulated spectrum and the measured one. Finally, Fig.~\ref{fig:SR}(c) shows the pulse profile and the energy spread at the end of the last radiator. We see that the radiation was emitted by a small portion of the head. The produced pulse was therefore a short spike with a peak power of 2 GW and a pulse duration of 5 fs. The Spectrum profiles, both in simulation and in measurement Fig.~\ref{fig:SR}(f) and (l), showed a single pulse with some internal modulation but sidebands did not appear.

The direct autocorrelation measurement of the FEL pulse duration was carried out at the low-density matter (LDM) beamline.
Autocorrelation in the EUV or soft-X-ray regime is more difficult and less flexible
than cross-correlation, but for such short pulses, it is the most reasonable option. The measurement was realized by splitting and recombining
the FEL pulse with the FERMI split-and-delay line autocorrelator/delay-creator and by monitoring the (energy-resolved) signal of the two-photon above-threshold ionization of argon.

\section{summary and conclusion}
 Well-controlled X-ray pulses with durations in the few-femtosecond and even
the sub-femtosecond range is essential in the investigation of dynamical processes at the time scale of the motion of
bound electrons in chemical, physical and biological studies.
Since soft-X-ray wavelengths can access core
electrons of extremely important light elements such as carbon, nitrogen, and oxygen, fast soft-X-ray pulses can drive the system under study into a regime where the transient excitation
is not depleted by competing fast channels. 
Novel free-electron lasers, producing ultrashort pulses with high peak power from the
extreme ultraviolet (EUV) to the soft-X-ray region have opened up a wide
range of new opportunities for scientists. 
 In this review paper, we described the progress and results of recent works on ultrafast pulse generation and measurement of seed laser-assisted FEL pulse in the range of EUV to soft X-Ray. We reported on the implementation of different approaches in the FERMI facility. The pulse duration of the seeded FEL depends on the seed laser pulse and longitudinal dispersion parameters, determining the bunching factor at the beginning of the FEL radiator. Theoretically from the bunching factor we can estimate the FEL pulse duration. 
The FERMI team expressed the cross-correlation, autocorrelation, and SPIDER approaches at seeded FEL and articulated the advantages and disadvantages of each method.
Basically, proper control of the seed laser pulse and
electron beam parameters governs the phase profile in the FEL pulse.
By imposing a proper chirp on seed laser frequency, we are able to shape the FEL pulse in both temporal and spectral domains in either HGHG or EEHG performances. The implementation of the chirp pulse amplification technique at seed FEL in the range of XUV showed the three times compression and gave us hope that this technique can be extended into the soft x-ray spectral window reducing the pulse
duration even below the femtosecond, with a very low transmission efficiency (0.2$\%$). The pulse duration measurement in this experiment was executed by a cross-correlation approach assisted by photo-emission of Helium gas.
Furthermore, we articulated the generation of a superradiant spike with the nonlinear evolution approach. In this approach, we considered the cascade superradiant process, and by employing the lower harmonic propagation, a small portion of the electron beam was manipulated for nonlinear FEL amplification. The FERMI team was able to generate a 5 fs FWHM pulse duration in 14.7 nm wavelength via this technique. The pulse duration measurement in this experiment was carried out by the autocorrelation method associated with the photo-emission of Argon gas. For seeded FEL applications, all presented techniques promise new possibilities for the FEL user community.    
\section{acknowledgments}
 The author expresses her sincere appreciation and gratitude to the FERMI team for their exceptional contributions and pioneering work on short pulse generation and superradiance experiments conducted at FERMI\cite{Mirian2021}. The author would like to extend special acknowledgment for the fruitful discussions with Luca Giannessi and Enrico Allaria.   
\bibliography{main}
\bibliographystyle{spiebib} 
\end{document}